\documentclass[fleqn,usenatbib]{mnras}

\usepackage{newtxtext,newtxmath}

\usepackage[T1]{fontenc}

\DeclareRobustCommand{\VAN}[3]{#2}
\let\VANthebibliography\thebibliography
\def\thebibliography{\DeclareRobustCommand{\VAN}[3]{##3}\VANthebibliography}

\usepackage{graphicx}	
\usepackage{amsmath}	
\usepackage{multirow}
\usepackage{array}
\usepackage{caption}

\definecolor{orange}{rgb}{1.0,0.5,0.}

\def\MDM{\ifmmode{\>M_{\textnormal{\sc dm}}}\else{$$M_{\textnormal{\sc dm}}}\fi}

\def\XH{\ifmmode{\>X_{\textnormal{\sc h}}} \else{$X_{\textnormal{\sc h}}$}\fi}
\def\nH{\ifmmode{\>n_{\textnormal{\sc h}}} \else{$n_{\textnormal{\sc h}}$}\fi}

\def\maspyr{\ifmmode{\>\textnormal{mas~yr}^{-1}}\else{mas~yr$^{-1}$}\fi}

\def\mG{\ifmmode{\>\mu\mathrm{G}}\else{$\mu$G}\fi}
\def\erg{\ifmmode{\> {\rm erg}}\else{erg}\fi}
\def\keV{\ifmmode{\> {\rm keV}}\else{keV}\fi}

\def\deg{\ifmmode{\>^{\circ}}\else{$^{\circ}$}\fi}
\def\onedeg{\ifmmode{\>1^{\circ}}\else{$1^{\circ}$}\fi}

\def\xvir{\ifmmode{\>\!x_{vir}}\else{$x_{vir}$}\fi}
\def\Mvir{\ifmmode{\>\!M_{vir} }\else{$M_{vir} $}\fi}
\def\rvir{\ifmmode{\>\!r_{vir}}\else{$r_{vir}$}\fi}
\def\vvir{\ifmmode{\>\!v_{vir}}\else{$v_{vir}$}\fi}
\def\Vvir{\ifmmode{\>\!V_{vir} }\else{$V_{vir} $}\fi}

\def\tratio{\ifmmode{\>\tau}\else{$\tau$}\fi}

\def\rms{\ifmmode{\>r_{\textnormal{\sc ms}}}\else{$r_{\textnormal{\sc ms}}$}\fi}

\def\Mpc{\ifmmode{\>\!{\rm Mpc}} \else{Mpc}\fi}
\def\kpc{\ifmmode{\>\!{\rm kpc}} \else{kpc}\fi}
\def\pc{\ifmmode{\>\!{\rm pc}} \else{pc}\fi}

\def\Gyr{\ifmmode{\>\!{\rm Gyr}} \else{Gyr}\fi}
\def\Myr{\ifmmode{\>\!{\rm Myr}} \else{Myr}\fi}
\def\yr{\ifmmode{\>\!{\rm yr}} \else{yr}\fi}
\def\pyr{\ifmmode{\>\!{\rm yr}^{-1}}\else{yr $^{-1}$} \fi}
\def\s{\ifmmode{\>\!{\rm s}}\else{s}\fi}
\def\ps{\ifmmode{\>\!{\rm s}^{-1}}\else{s$^{-1}$}\fi}
\def\Hz{\ifmmode{\>\!{\rm Hz}}\else{Hz}\fi}

\def\kms{\ifmmode{\>\!{\rm km\,s}^{-1}}\else{km~s$^{-1}$}\fi}

\def\K{\ifmmode{\>\!{\rm K}}\else{K}\fi}

\def\sr{\ifmmode{\>\!{\rm sr}}\else{sr}\fi}
\def\psr{\ifmmode{\>\!{\rm sr}^{-1}}\else{sr$^{-1}$}\fi}
\def\arcs{\ifmmode{\>\!{\rm arcsec}}\else{arcsec}\fi}
\def\parcs{\ifmmode{\>\!{\rm arcsec}^{-1}}\else{arcsec${-1}$}\fi}
\def\parcss{\ifmmode{\>\!{\rm arcsec}^{-2}}\else{arcsec${-2}$}\fi}

\def\cm{\ifmmode{\>\!{\rm cm}}\else{cm}\fi}
\def\cc{\ifmmode{\>\!{\rm cm}^{3}}\else{cm$^{3}$}\fi}
\def\sqc{\ifmmode{\>\!{\rm cm}^{2}}\else{cm$^{2}$}\fi}
\def\pcc{\ifmmode{\>\!{\rm cm}^{-3}}\else{cm$^{-3}$}\fi}
\def\psc{\ifmmode{\>\!{\rm cm}^{-2}}\else{cm$^{-2}$}\fi}

\def\g{\ifmmode{\>\!{\rm g}}\else{g}\fi}
\def\Msun{\ifmmode{\>\!{\rm M}_{\odot}}\else{M$_{\odot}$}\fi}
\def\hMsun{\ifmmode{\> h^{-1}{\rm M}_{\odot}}\else{$h^{-1}$M$_{\odot}$}\fi}

\def\Zsun{\ifmmode{\>\!{\rm Z}_{\odot}}\else{Z$_{\odot}$}\fi}

\def\Lsun{\ifmmode{\>\!{\rm L}_{\odot}}\else{L$_{\odot}$}\fi}

\def\rayl{\ifmmode{\>\!{\rm R}}\else{R}\fi}
\def\mR{\ifmmode{\>\!{\rm mR}}\else{mR}\fi}

\renewcommand{\ion}[2]{\hbox{#1\,{\sc #2}}}

\def\lya{\ifmmode{\>\!{\rm Ly}\alpha}\else{Ly$\alpha$}\fi}

\def\Ha{\ifmmode{\>\!{\rm H}\alpha}\else{H$\alpha$}\fi}
\def\Hb{\ifmmode{\>\!{\rm H}\beta}\else{H$\beta$}\fi}

\def\HI{\ifmmode{\> \textnormal{\ion{H}{i}}} \else{\ion{H}{i}}\fi}
\def\HII{\ifmmode{\> \textnormal{\ion{H}{ii}}} \else{\ion{H}{ii}}\fi}
\def\CIV{\ifmmode{\> \textnormal{\ion{C}{iv}}} \else{\ion{C}{iv}}\fi}
\def\SiIV{\ifmmode{\> \textnormal{\ion{S}{iv}}} \else{\ion{Si}{iv}}\fi}

\def\NH{\ifmmode{\> {\rm N}_{\rm H}} \else{N$_{\rm H}$}\fi}
\def\Ng{\ifmmode{\> {\rm N}_{\rm gas}} \else{N$_{\rm gas}$}\fi}
\def\NHI{\ifmmode{\> {\rm N}_{\HI}} \else{N$_{\HI}$}\fi}
\def\MHI{\ifmmode{\> {\rm M}_{ \HI}} \else{M$_{\HI}$}\fi}

\def\mua{\ifmmode{\>\mu_{ \textnormal{\Ha}}}\else{$\mu_{ \textnormal{\Ha}}$}\fi}
\def\alphabha{\ifmmode{\>\alpha_{B}^{(\textnormal{\Ha})}}\else{$\alpha_{B}^{(\textnormal{\Ha})}$}\fi}

\title[Turbulence driving in galaxies]{The driving mode of turbulence in disc galaxy simulations with adaptive mesh refinement}

\author[Watt et al.]{
James Watt,$^{1}$\thanks{E-mail: James.Watt@anu.edu.au (JW)}
Christoph Federrath,$^{1}$
Thor Tepper-Garcia,$^{2}$
Oscar Agertz$^{3}$
and Joss Bland-Hawthorn$^{2}$
\\
$^{1}$Research School of Astronomy and Astrophysics, Australian National University, Cotter Road, Canberra, ACT 2611, Australia\\
$^{2}$Sydney Institute for Astronomy, School of Physics, A28, The University of Sydney, NSW 2006, Australia\\
$^{3}$Lund Observatory, Division of Astrophysics, Department of Physics, Lund University, Box 118, SE-221 00 Lund, Sweden
}

\date{Accepted XXX. Received YYY; in original form ZZZ}

\pubyear{\the\year{}}

\begin{document}
\label{firstpage}
\pagerange{\pageref{firstpage}--\pageref{lastpage}}
\maketitle

\begin{abstract}
Turbulence is a key ingredient in controlling the structure of the interstellar medium (ISM) and the formation of stars. However, we still lack a detailed understanding of the drivers of turbulence in the ISM of galaxies. Previous idealised simulations of turbulence employ a stochastic forcing field to drive turbulence. The geometry of this forcing field — whether it is predominantly solenoidal or compressive — is a key parameter governing how turbulence shapes the ISM density distribution and regulates the star formation rate. This is commonly expressed through the turbulence driving parameter ($b$), which quantifies the relative contribution of compressive versus solenoidal driving. Therefore, an accurate knowledge of the driving parameter is essential for understanding and predicting star formation, and for sub-grid modelling of ISM physics and the star formation rate. In this work, we introduce an algorithm to measure the turbulence driving parameter in adaptive mesh refinement (AMR) simulations of galaxies on variable turbulence kernel sizes. We focus our analysis on a synthetic Large Mallenanic Cloud (LMC), present-day analogue with a total mass $M \approx 10^{11}$\,\Msun. We find that turbulence is driven primarily solenoidally ($b<0.4$) within the inner $\sim4\,\kpc$ of the galaxy, and becomes increasingly compressive with $b>0.4$ towards the outskirts, $R\gtrsim4.5\,\kpc$. The volume-weighted median across the disc, $b\simeq0.4$, is consistent with the natural mixture of driving modes. We further find that $b$ is weakly correlated with the strength of shear, in that solenoidal driving tends to be associated with regions of higher shear, as expected for the more central parts of galaxies. These trends persist over $\sim\!2$\,Gyr of the galaxy’s evolution, and are largely insensitive to the choice of turbulence kernel size.
\end{abstract}




\section{Introduction}

Turbulence in the interstellar medium (ISM) plays a central role in regulating the evolution of galaxies. The driving mechanism of turbulence is a key parameter in determining the density distribution of gas \citep{MolinaEtAl2012, Hopkins2013, NolanEtAl2015}, setting the phase fraction in the multiphase ISM \citep{SeifriedEtAl2011, MicicEtAl2012} and regulating star formation \citep{FederrathKlessen2012, PadoanEtAl2014, MathewFederrathSeta2023}. A key diagnostic of the turbulent forcing is the turbulence driving parameter $b$, defined by
\begin{equation}
    b=\frac{\sigma_{\rho/\rho_0}}{\mathcal{M}}\,,
    \label{eq:density_variance_mach_relation}
\end{equation}
where $\sigma_{\rho/\rho_0}$ is the standard deviation of the ratio of density to the mean density and $\mathcal{M}$ is the root-mean-square (rms) Mach number. The $b$ parameter has been used to quantify the relative contribution of solenoidal (divergence-free) and compressive (curl-free) driving modes in periodic-box simulations of stochastically driven turbulence in non-magnetized isothermal gas \citep[see][]{PadaonEtAl1997, PassotSemadeni1998, FederrathEtAl2008, FederrathEtAl2010, PriceFederrath2010, KonstandinEtAl2012ApJ}, as well as magnetized isothermal and non-isothermal plasmas \citep{MolinaEtAl2012, FederrathBanerjee2015, NolanEtAl2015}. Purely solenoidal driving, which can be thought of as turbulence driven by a divergence free stochastic forcing field, corresponds to $b\approx1/3$, while purely compressive driving yields $b\approx1$, with mixed driving producing intermediate values. The $b$ parameter is especially relevant for star formation studies as varying $b$ can change star formation rates by an order of magnitude \citep[see][]{FederrathKlessen2012, SalimEtAl2015, Federrath2018} and changes the characteristic mass of the IMF \citep{MathewFederrathSeta2023}.

Understanding the spatial variation of $b$ within galaxies and molecular clouds offers a unique window into the physical processes responsible for stirring the ISM and regulating star formation. Some work has been done in this direction, particularly focusing on measuring $b$ in selective regions of various galaxies. For instance, \citet{GinsburgEtAl2013} measured the turbulence driving parameter of the system dubbed GRSMC 43.30-0.33, \citet{FederrathEtAl2016} looked at the central molecular zone cloud G0.253+0.016, \citet{MenonEtAl2020} measured turbulence driven by ionizing feedback in the pillars of the Carina Nebula, \citet{ShardaEtAl2022} measured the $b$ parameter of the star-forming region N159E in the Large Magellanic Cloud, \citet{MarchalEtAl2021} looked at the driving of turbulence in the high latitude WNM of the Milky Way, \citet{IsabellaEtAl2023, GerrardEtAl2024} have produced maps of the $b$ parameter using HI observations of the Small Magellanic Cloud and high latitude clouds in the Milky Way. Finally, \citet{MillerEtAl2026} is the only observational work so far that as studied $b$ in full-disc observations (in NCG1313 and NGC7793). The driving mechanism of turbulence has also been studied in controlled simulations of molecular clouds \citep{KortgenEtAl2017}, idealized box simulations of the ISM \citep{SeifriedEtAl2011}, and kpc-sized shearing boxes of galactic discs \citep{GerrardFederrath2026}. The only study of turbulence driving in full-scale galaxy simulations to date is that presented by \citet{JinEtAl2017}, however, resolution constraints did not allow them for a robust determination of $b$. In the realm of numerical simulations more broadly, works like \citet{FederrathEtAl2011}, \citet{PanEtAl2016}, and \citet{GrisdaleEtAl2017} have quantified the fraction of kinetic energy in compressive vs.~solenoidal motions in the gas. Similarly, such characterisation has been done in observations \citep{OrkiszEtAl2017}. However, the ratio of solenoidal to compressible modes in the velocity field is not a direct probe of the $b$ parameter relevant for determining star formation efficiency, ISM phase, and density distribution of the gas. In fact, even purely compressive driving induces solenoidal motions in the gas at a level as high as $50\%$ of the total kinetic energy \citep{FederrathEtAl2010}. 

In this work, we extend the algorithm for measuring the $b$ parameter introduced in \citet{IsabellaEtAl2023} to adaptive-mesh refinement (AMR) simulations. We use it to study the driving of turbulence in a present-day Large Magellanic Cloud (LMC) analogue with an initial gas fraction of $20\%$, created with the {\sc Nexus} simulation framework \citep{GarciaEtAl2024}.\\

The rest of this paper is structured as follows: in Section~\ref{sec:galaxy_description}, we briefly describe our galaxy simulations. Section~\ref{sec:method} introduces the algorithm for calculating $b$. Section~\ref{sec:results} demonstrates how the driving parameter of turbulence varies across the simulated galaxy in terms of radius, scale height, and over time, and how it is correlated with shear. Finally, we summarise our results in Section~\ref{sec:conclusions}.

\section{Galaxy simulation}
\label{sec:galaxy_description}

The main goal of this work is to study the driving of turbulence over galactic scales, rather than just for individual clouds. We choose to do so using a system that resembles today's LMC. Thus, we focus our attention on a single galaxy model, and defer a comparative analysis of different galaxy models to future work.

\subsection{Initial conditions}

We use one simulation of the suite introduced by and \citet[][see also \citealt{bla25a}]{bla24a}. Specifically, we adopt the high-resolution version of synthetic galaxy dubbed `fd50\_fg00\_nac', a multi-component system with a dark matter (DM) halo mass of $10^{11}\,$\Msun, a total disc mass of $10^{10}\,$\Msun, of which 20\% is initially in the form of gas \citep[see][tab.~1]{bla24a}. 

The galaxy models are constructed and evolved with the {\sc Nexus} framework \citep{GarciaEtAl2024}. Particle positions and velocities of each of the components making up the galaxy are generated with the self-consistent modelling module (SCM) provided by the Action-based GAlaxy Modelling Architecture ({\sc agama}) stellar dynamical library \citep{Vasiliev2019}, which has been complemented to treat gas in addition to the DM and stellar components. The DM halo virial radius and scale radius are $R_{\rm vir} = 37$\,\kpc, and $r_s = 9.2$\,\kpc, respectively.
The total mass and structural parameters of the halo are broadly consistent with today's LMC \citep[e.g.][]{vas21x}.\footnote{This galaxy model is also broadly consistent with a Milky-Way progenitor at $z \approx 3$ \citep[][see also \citealt{bla24a}]{bla16a}.}
The pre-assembled stellar disc and the gas disc both have an initial scale length $R_d \approx 1.8$\,\kpc. The initial scale height of the stellar disc is roughly $z_d \approx 0.2$\,\kpc; the initial scale height of the gas disc is roughly $10\times$ less than that at the centre, and increases with radius (`flaring' disc). It is worth mentioning though that its initial value is irrelevant, as the disc structure necessarily (and significantly) changes once the disc is allowed to undergo cooling and heating, and to form stars.

The initial conditions are evolved with the adaptive mesh refinement (AMR), $N$-body/hydrodynamical code {\sc Ramses} \citep{Teyssier2002}, augmented with the galaxy formation physics (incl. star formation and feedback processes) outline in \citet[][]{age13} and \citet[][]{age21l}, see \citet{bla24a} for more details on the specific setup in this work.

The DM halo, pre-assembled stellar disc, and gas disc, are sampled, respectively, with $2\times10^6$, $2\times10^6$, and $10^6$ particles. The latter number is only relevant for the construction of the initial conditions -- at runtime, the gas particles are mapped onto the computational grid of {\sc Ramses}, resulting in $\sim3\times10^6$ cells, which changes during the evolution of the system as a consequence of the adaptive nature of the grid (see below). The composite system is evolved in a cubic simulation volume with a length of 100\,kpc per side. The total simulation timespan is about 2\,Gyr. At runtime, the AMR grid is maximally refined up to level~14, implying a limiting spatial resolution of \mbox{100\,kpc / $2^{14} \approx 6\,$pc}.\footnote{Note that this marginally resolves the scale height of the gas disc close to the centre at initialisation. However, as explained above, it is irrelevant since the disc 1) has a larger scale height away from the centre, and 2) will thicken quickly and significantly at runtime as a result of stellar feedback.}
To allow the gas disc to settle, we evolve the system with an adiabatic equation of state ($\gamma = 5/3$, appropriate for a monoatomic gas) for 80\,Myr, after which cooling/heating, star formation and feedback are `switched on'.
We refer the reader to \citet{GarciaEtAl2024} for further details about the initial conditions and the simulation techniques -- and, in particular, \citet{bla24a,bla25a} for the galaxy formation module.

\subsection{Galaxy properties and morphology}

\begin{figure*}
    \centering
    \includegraphics[width=1\linewidth]{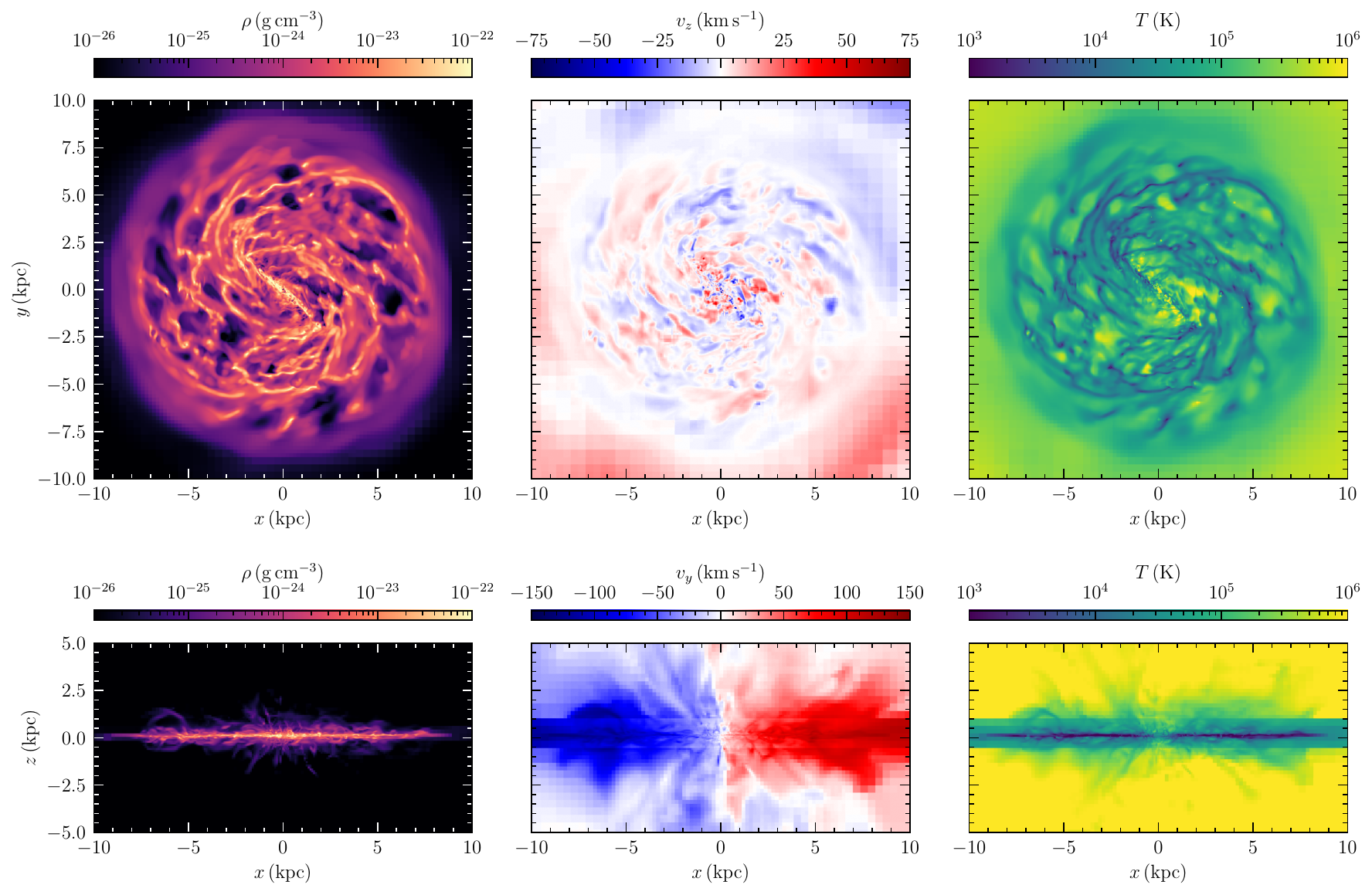}
    \caption{Density-weighted projections of the gas density (left column), line-of-sight velocity (middle column) and temperature (right column) after $\sim 2\,\mathrm{Gyr}$ of evolution. The interaction of the dark matter, stellar, and gas components, galactic dynamics, heating/cooling, and star formation feedback, have created a rich multi-phase structure with spiral features, clouds, filaments, and a central bar. This is the final state of the galaxy used to analyse the turbulence properties in the disc, with the main focus on the turbulence driving parameter $b$ (cf.~Eq.~\ref{eq:density_variance_mach_relation}). However, we also study the properties at several times in the earlier evolution, below.}
    \label{fig:galaxy_projections}
\end{figure*}

Figure~\ref{fig:galaxy_projections} shows the state of the gas at the end of the simulation ($t \approx 2$\,Gyr). The top and bottom panels in the left column show the projection of the gas density along the $z$- and $y$-axes, respectively. The galaxy has developed sub-structures and flocculent spiral arms. The top panel in the middle column displays the density-weighted value of the gas' vertical velocity $v_z$ projected along the $z$-axis; and similar for the $v_y$ component in the bottom panel. In this projection, the disc's rotation is discernible (blue-/redshift). The right column displays the density-weighted projection of the temperature along the $z$-axis (top) and the $y$-axis (bottom). As a result of stellar feedback, and the interplay between cooling and heating, the gas disc features diffuse pockets of hot material entrained within dense filamentary knots close to the plane, and plumes of gas ejected from the plane.

\section{Turbulence analysis method}
\label{sec:method}

Equation~\eqref{eq:density_variance_mach_relation} is based on idealised simulations of driven turbulence where the velocity and density fluctuations are induced entirely by turbulent motions. However, in real galaxies, the density and velocity distributions contain large-scale, systematic features and flows that originate from the inherent structure of a galaxy, such as large-scale rotation, gravitational gradients, etc. Not all of these motions and structures are turbulent. Therefore, in order to isolate the turbulent fluctuations, large-scale, systematic features must be subtracted before flow variables can be used for calculating $b$ \citep{FederrathEtAl2016, StewartFederrath2022, NarayanTritsisFederrath2025}. Here, we explain our algorithm for extracting the turbulent components of the flow from the synthetic galaxy's ISM, with the ultimate goal of calculating $b$.

\subsection{Roving and smoothing kernel}

Consider a set $\mathcal{U}$ of all the cell-centred coordinates $[x, y, z]$ in an AMR-based galaxy simulation. Suppose $\mathcal{P}$ is the collection of points in $\mathcal{U}$ that satisfy some criteria set on the flow variables, for instance, a temperature criterion to restrict the analysis to just one gas phase, say the warm neutral medium. We define the functions
\begin{align}
    &D: \mathcal{P} \rightarrow \mathbb{R}^3, \quad D(p)=[dx, dy, dz]\,,\\
    &V: \mathcal{P} \rightarrow \mathbb{R}, \quad V(p)=dx\,dy\,dz\,,
\end{align}
where $dx$, $dy$ and $dz$ are the widths of the cells. Suppose we want to calculate the turbulence driving parameter at $p\in\mathcal{P}$. Then, we define a roving kernel centred at $p$ as 
\begin{equation}
    \mathcal{R}_{p}=\{a\in\mathcal{P}:||(p-a)||^2\le\alpha||D(p)||^2\}\,,
    \label{eq:Rp_def}
\end{equation}
where $||\cdot||$ is the Euclidean norm defined by $||[x, y, z]||=\sqrt{x^2+y^2+z^2}$. The roving kernel is a Gaussian of full width at half maximum $\lambda||D(p)||$, truncated at a radius of $1.3$~times that FWHM (i.e.\ at $3\sigma$), so that $\alpha=1.3\lambda$. Throughout this work we adopt $\lambda=5$, and hence $\alpha=6.5$; this choice is justified in Appendix~\ref{app:kernel_size}.

For $a\in\mathcal{R}_{p}$, define a smoothing kernel $\mathcal{S}_{a|p}$ centred at $a$ as
\begin{equation}
    \mathcal{S}_{a|p}=\{c\in\mathcal{P}:||c-a||^2\le\alpha||D(p)||^2/4\}\,.
    \label{eq:Sa_def}
\end{equation}
Note that the extent of the smoothing kernel is set by the cell size at the centre of the roving kernel, $||D(p)||$, rather than by that of the cell $a$ at which the smoothing is evaluated. A single smoothing scale is therefore applied uniformly across $\mathcal{R}_{p}$, so that the background subtracted at every point of a given roving kernel is defined with respect to the same physical length scale. Consequently, the smooth component depends on both $a$ and $p$, as reflected in the notation below.
Suppose $Q:\mathcal{U}\rightarrow\mathbb{R}$ is a flow variable. Then, we can calculate the Gaussian weighted mean of $Q$ over a kernel $\mathcal{K}_{c}$ centred at $c$ as 
\begin{equation}
    \langle Q\rangle_{(\mathcal{K}_{c};\,\mathrm{FWHM})} = \frac{\sum\limits_{a\in\mathcal{K}_{c}}Q(a)V(a)\exp{\left(-\frac{||c-a||^2}{2\sigma^2}\right)}}{\sum\limits_{a\in\mathcal{K}_{c}}V(a)\exp{\left(-\frac{||c-a||^2}{2\sigma^2}\right)}}\,,
    \label{eq:unstructured_gaussian_mean}
\end{equation}
where $\sigma=\mathrm{FWHM}/2.355$. For $a\in\mathcal{R}_{p}$, we define the smooth background component of $Q$ at $a$ for the kernel centred at $p$ as 
\begin{equation}
    Q_{\mathrm{S}|p}(a)=\langle Q\rangle_{(\mathcal{S}_{a|p};\,\beta||D(p)||)}\,,
    \label{eq:smoothing}
\end{equation}
where $\beta=\lambda/2=2.5$, i.e.\ the smoothing kernel has half the FWHM of the roving kernel. Then, the turbulent component of $Q$ at point $a$ for the kernel centred at $p$ is given by
\begin{equation}
    Q_{\mathrm{T}|p}(a)=Q(a)-Q_{\mathrm{S}|p}(a).
    \label{eq:turb_component_calculation}
\end{equation}

\subsection{Density dispersion and Mach number}

We now describe the procedure for calculating $\sigma_{\mathrm{\rho/\rho_0}}$ and $\mathcal{M}$. We need to subtract the large-scale density structures and bulk velocities. We note that \citet{IsabellaEtAl2023} perform this subtraction on the line-of-sight velocity component and the logarithm of density ($\ln\rho$). The reason behind this seemingly odd choice of $\ln\rho$ is that while turbulent velocity fluctuations are additive, turbulent density fluctuations are multiplicative \citep[see][]{Semadeni1994}, making $\ln\rho$ the natural variable for obtaining the turbulent component. 

For $a\in\mathcal{R}_{p}$, we define 
\begin{equation}
    \left(\frac{\rho}{\rho_0}\right)_\mathrm{turb}(a) = \frac{\exp({\ln\rho_{\mathrm{T}|p}(a))}}{\sum\limits_{b\in\mathcal{R}_{p}}\exp({\ln\rho_{\mathrm{T}|p}(b))}}\,,
\end{equation}
and, for each velocity component $i\in\{\mathrm{x},\mathrm{y},\mathrm{z}\}$,
\begin{equation}
    M_{\mathrm{turb},\,i}(a)=\frac{v_{\mathrm{i}_{\mathrm{T}|p}}(a)}{c_\mathrm{s}(a)}\,.
    \label{eq:component_mach}
\end{equation}

Then, for $p\in\mathcal{P}$, we define
\begin{equation}
    \sigma_\mathrm{\rho/\rho_0}(p)=\sqrt{\left\langle\left(\frac{\rho}{\rho_0}\right)_\mathrm{turb}^2\right\rangle_{(\mathcal{R}_{p};\,\gamma)} - \left\langle\left(\frac{\rho}{\rho_0}\right)_\mathrm{turb}\right\rangle_{(\mathcal{R}_{p};\,\gamma)}^2}\,,
    \label{eq:sigma_rho}
\end{equation}
\begin{equation}
    \mathcal{M}(p)=\sqrt{\sum_{i\in\{\mathrm{x},\mathrm{y},\mathrm{z}\}}\sigma^2_{M_{\mathrm{turb},\,i}}(p)}\,,
    \label{eq:mach_calculation}
\end{equation}
with the dispersion of each component given by
\begin{equation}
    \sigma_{M_{\mathrm{turb},\,i}}(p)=\sqrt{\left\langle M^2_{\mathrm{turb},\,i}\right\rangle_{(\mathcal{R}_{p};\,\gamma)} - \left\langle M_{\mathrm{turb},\,i}\right\rangle^2_{(\mathcal{R}_{p};\,\gamma)}}\,.
\end{equation}
In both cases $\gamma=\lambda||D(p)||$, i.e.\ the moments are taken with the same Gaussian weight that defines the roving kernel. The turbulence driving parameter $b(p)$ is then calculated as per Eq.~(\ref{eq:density_variance_mach_relation}), for point $p$,
\begin{equation}
    b(p)=\frac{\sigma_{\rho/\rho_0}(p)}{\mathcal{M}(p)}.
\end{equation}
Following \citet{ShardaEtAl2022}, we reject results for any $p\in\mathcal{P}$ if $\mathcal{R}_{p}$ has 30 or fewer cells.

\subsection{Choice of $\alpha$, $\beta$, and $\gamma$}

While the notation looks complicated, the crux of the algorithm is similar to that of \citet{IsabellaEtAl2023} and \citet{MillerEtAl2026}, who have also shown that the measurement of $b$ is largely independent of the kernel size, as long as the kernel size is not varied by factors of several, in which case the turbulence on a different kernel scale is probed. The relative insensitivity of $b$ with respect to changes in the kernel size arises because, while both $\sigma_\rho$ and $\mathcal{M}$ are strongly scale-dependent, the ratio of the two, i.e., $b$, is only weakly dependent on kernel size (see Appendix~\ref{app:kernel_size}). Therefore, we choose $\alpha$, so as to have a sufficiently large number of cells in each kernel, while at the same time keeping it as small as possible to allow for an exploration of the spatial variation of the turbulence quantities across the galaxy.

We construct a roving kernel ($\mathcal{R}_{p}$) around a cell ($p$) in the simulation data. As defined above, its FWHM is $\lambda||D(p)||$, i.e.\ $\lambda=5$ times the diagonal $||D(p)||=\sqrt{dx^2+dy^2+dz^2}$ of the cell $p$, and it is truncated at $1.3$~FWHM ($3\sigma$), giving $\alpha=1.3\lambda=6.5$. Note that the size of the roving kernel is adaptive unlike in \citet{IsabellaEtAl2023} and \citet{MillerEtAl2026}, to account for the AMR nature of our datasets.

Eq.~\eqref{eq:smoothing} evaluates the smooth background $Q_{\mathrm{S}|p}$ of a flow variable $Q$ at every cell of the roving kernel, by convolving $Q$ with a second Gaussian --- the smoothing kernel --- of FWHM $\beta||D(p)||$ with $\beta=\lambda/2=2.5$, i.e.\ half the FWHM of the roving kernel. The smoothing kernel is likewise truncated at $3\sigma$. Defining it using Eq.~\eqref{eq:unstructured_gaussian_mean} has the advantage of the volume weighting taking into account the AMR nature of the kernel and the normalization, allowing for gaps in the kernel in case some cells do not satisfy the physical criteria (like the choice of the ISM phase) used for constructing $\mathcal{P}$. Eq.~\eqref{eq:turb_component_calculation} calculates the turbulent component of the flow by subtracting the Gauss-smoothed background. This procedure removes all flow features that appear at length scales greater than half the size of the roving kernel. The turbulent Mach number and the density dispersion are calculated over a roving kernel with the same Gauss weighting as in Eq.~\eqref{eq:unstructured_gaussian_mean}. 

\subsection{Choice of ISM phase}
\label{sec:galaxy_description_WNM}

The ISM is composed of multiple thermally distinct phases. The molecular phase consists predominantly of molecular hydrogen ($\mathrm{H}_2$) and is generally the coldest and densest phase, with typical temperatures of $T\sim10-30,\mathrm{K}$. The cold neutral medium (CNM) consists of dense, atomic gas and is typically characterized by $30,\mathrm{K}\le T\le200\,\mathrm{K}$. The warm neutral medium (WNM) is more diffuse and typically lies in the range $5000\,\mathrm{K}\le T\le8000\,\mathrm{K}$. The unstable neutral medium (UNM) occupies an intermediate regime of $200\,\mathrm{K}\le T\le5000\,\mathrm{K}$ between the CNM and WNM, where gas is thermally unstable to perturbations and can evolve towards either the cold or warm phase depending on local heating and cooling conditions. At higher temperatures ($T\ge10^4\,\mathrm{K}$) are the warm ionized medium (WIM) and the hot ionized medium (HIM), usually found in supernova-heated cavities. 

The multi-phase nature of the ISM can significantly increase the density dispersion. This can lead to an overestimate of the turbulent density dispersion, as multiple phases can coexist without significant motions between the phases \citep{MohapatraEtAl2022}. Furthermore, Eq.~\eqref{eq:density_variance_mach_relation} is based on studies of isothermal turbulence. The Mach number, which depends on the sound speed, is therefore also affected by changes in the temperature, and the presence of different phases would consequently affect the turbulent Mach number. Therefore, as in \citet{IsabellaEtAl2023}, we restrict our analysis to a narrow temperature range and study turbulence only in the WNM.

\subsection{Example of a single kernel}

\begin{figure*}[h]
    \centering
    \includegraphics[width=1\linewidth]{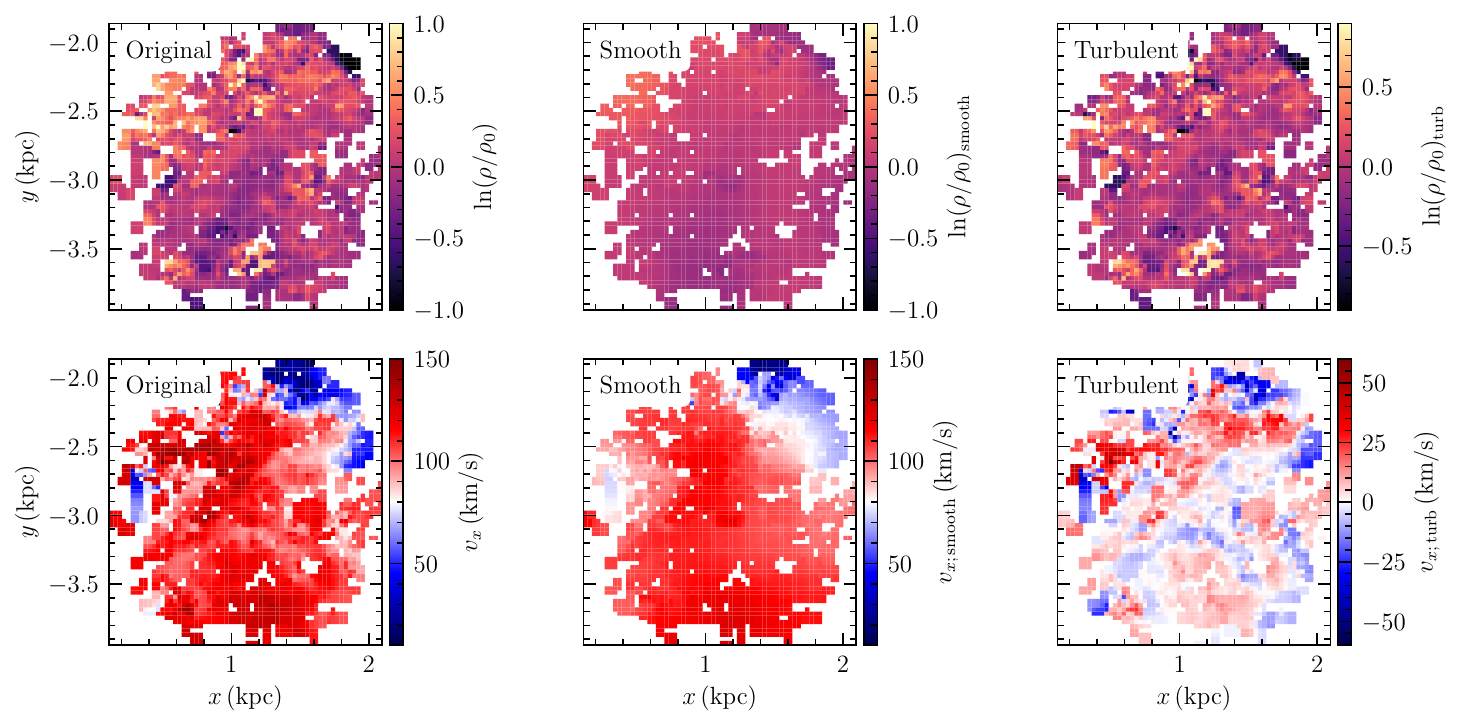}
    \caption{An example of a single roving kernel projected along the $z$-axis, illustrating the process of subtracting large-scale, non-turbulent structure from the density (top) and the $x$ component of the velocity (bottom). The axes are labelled following the same coordinates as in Figure~\ref{fig:galaxy_projections}. The left panels show the original fields: the natural logarithm of the normalized density (top) and the $x$ component of velocity (bottom) in a roving kernel truncated at 1.3~FWHM, where the FWHM is $\lambda=5$ times the diagonal $||D(p)||$ of the cell at the centre of the kernel. The middle panel shows the smooth component of $\ln(\rho/\rho_0)$ and $v_\mathrm{x}$. The last panel shows the turbulent fluctuations, which are clearly visible after the subtraction of the background.}
    \label{fig:kernel}
\end{figure*}

We now demonstrate the action of the algorithm on a single kernel for calculating $b$ at a point inside the galactic disc. We choose a WNM cell in the galaxy, close to the centre. Figure~\ref{fig:kernel} shows a single roving kernel (as defined in Eq.~\eqref{eq:Rp_def}) drawn around this cell. The kernel has a total radius of $1.3\times5\times\sqrt{\Delta x^2+\Delta y^2+\Delta z^2}$ (corresponding to 3$\sigma$ radius of the corresponding Gaussian and spanning $\approx 2\times2\,kpc$), where $\Delta x$, $\Delta y$ and $\Delta z$ are the cell sizes. The characteristic size of the kernel is set by the standard deviation of this Gaussian, $\sigma=0.35\,\kpc$. The top-left panels show a snapshot of the logarithm of the normalized density $\ln(\rho/\rho_0)$, in accordance with Eq.~\eqref{eq:unstructured_gaussian_mean}. The top-middle panel shows the smooth component of $\ln(\rho/\rho_0)$ calculated using Eq.~\eqref{eq:smoothing}. The smoothing was performed using a Gaussian with a FWHM half the size of the Gaussian corresponding to the roving kernel. One can see that small-scale features have been smoothed out yielding a map of the density structure originating from non-turbulent sources. The top-right panel shows the turbulent component calculated using Eq.~\eqref{eq:turb_component_calculation}. The bottom three panels demonstrate the same procedure for the $x$-component of the velocity. 

\begin{figure}
    \centering
    \includegraphics[width=1\linewidth]{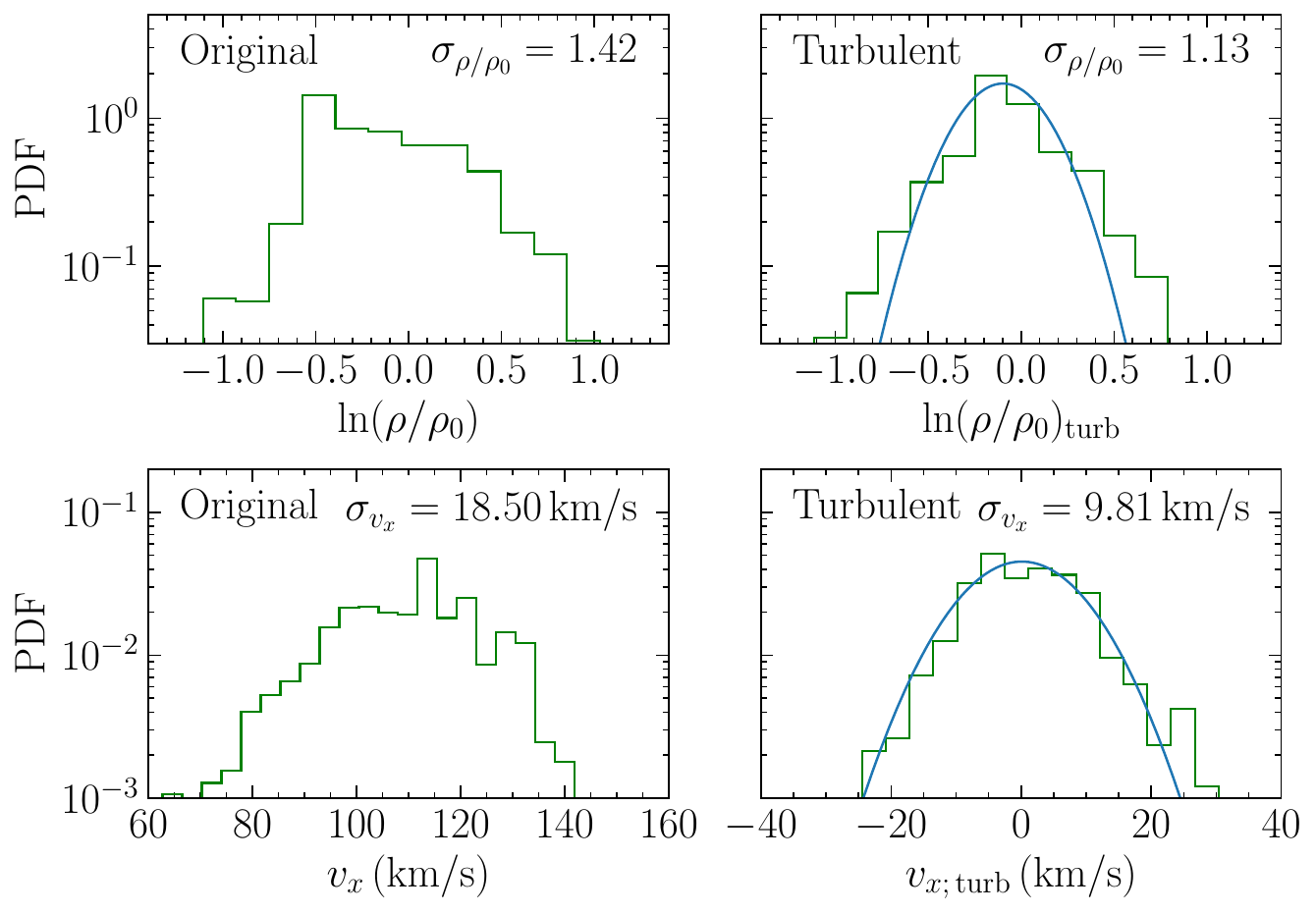}
    \caption{Top: The distribution of the log of the normalized density before (left panel) and after (right panel) background subtraction. The original data (left panel) has strong non-Gaussian components due to large-scale, non-turbulent contributions. Their subtraction reduces the density dispersion and yields a distribution close to a log-normal density distribution, a signature of turbulence. The deviation from a log-normal distribution towards the tails is due to intermittency or incomplete turbulence isolation, which however, only applies to $\approx 7\%$ of the data. Bottom: The distribution of the $x$ component of the velocity before (left panel) and after (right panel) background subtraction. Similar to the log of the normalized density, the original data (left panel) has strong non-Gaussian components originating from large-scale flows in the galaxy. Smoothing followed by subtraction yields a Gaussian distribution, indicating that we have isolated primarily turbulent fluctuations. Intermittency effects accounting for $\approx 2\%$ of the data can be seen in the non-Gaussian tails of the distribution.}
    \label{fig:den_vel_hist}
\end{figure}

The results of Figure~\ref{fig:kernel} can be understood more quantitatively by looking at the statistics of the density and the velocity. In isothermal turbulence, the density approximately follows a log-normal distribution \citep{Semadeni1994,PassotVazquez1998}, i.e., $\ln(\rho/\rho_0)$ is Gaussian, and the velocity components follow a normal distribution \citep[e.g.,][]{Federrath2013}. Figure~\ref{fig:den_vel_hist} shows a comparison between the density and the velocity statistics before and after background removal. The original data (left panels of Figure~\ref{fig:den_vel_hist}) have strong non-Gaussian components originating from large-scale structures and flows in the galactic disk. Background removal yields a density distribution close to a log-normal distribution and a nearly Gaussian velocity distribution (more specifically, the velocity components are Gaussian) for the major fraction (more than $90\%$) of the PDFs, both of which are strong signatures of turbulent motion \citep{KritsukEtAl2007,FederrathEtAl2010,Federrath2013}. The tails of the distributions deviate from the perfect log-normal and Gaussian distributions, which may be attributed to intermittency effects \citep[see][]{SheAndLeveque1994, KritsukEtAl2007, FederrathEtAl2010, KonstandinEtAl2012, Hopkins2013} as well as imperfect isolation of turbulent motions. However, these outliers constitute less than $10\%$ of the data and do not have a significant impact on the final measurements, i.e., the dispersions. 

\section{Results}
\label{sec:results}

\begin{figure*}
    \centering
    \includegraphics[width=1\linewidth]{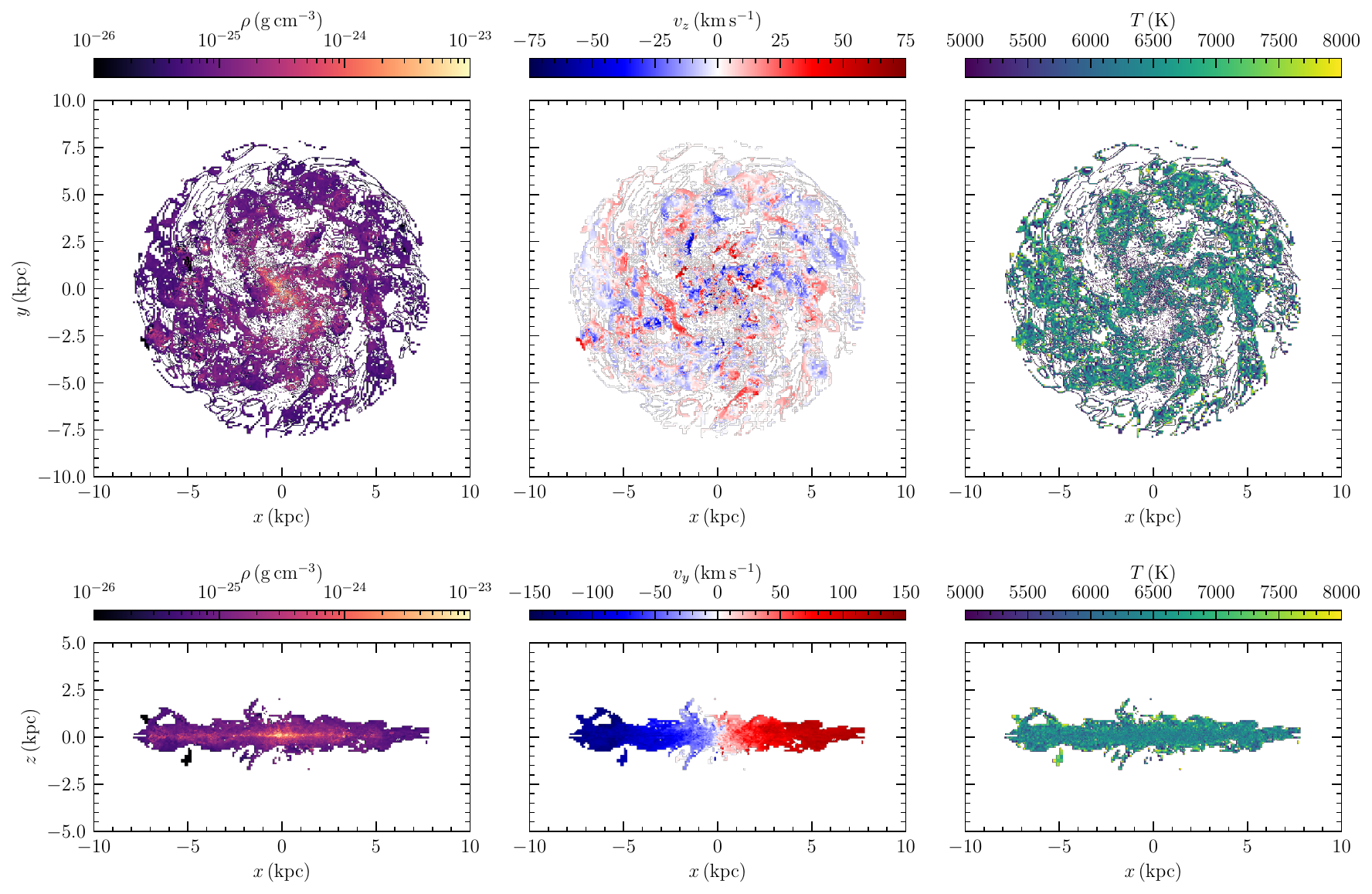}
    \caption{Same as Figure~\ref{fig:galaxy_projections}, but for WNM gas only.}
    \label{fig:wnm_projections}
\end{figure*}

In this section, we study the turbulence driving parameter in the WNM and determine its variation across the entire galactic disk. Figure~\ref{fig:wnm_projections} maps out the WNM in our galaxy snapshot, in order to focus on a single gas phase, required to robustly apply Eq.~(\ref{eq:density_variance_mach_relation}) -- cf.~Sec.~\ref{sec:galaxy_description_WNM}. While selection of the WNM phase obviously removes some of the gas, it still covers a substantial fraction of the galactic disc, allowing us to study the turbulence across the entire disc. 

\subsection{Variation across the galaxy}

\begin{figure*}
    \centering
    \includegraphics[width=1\linewidth]{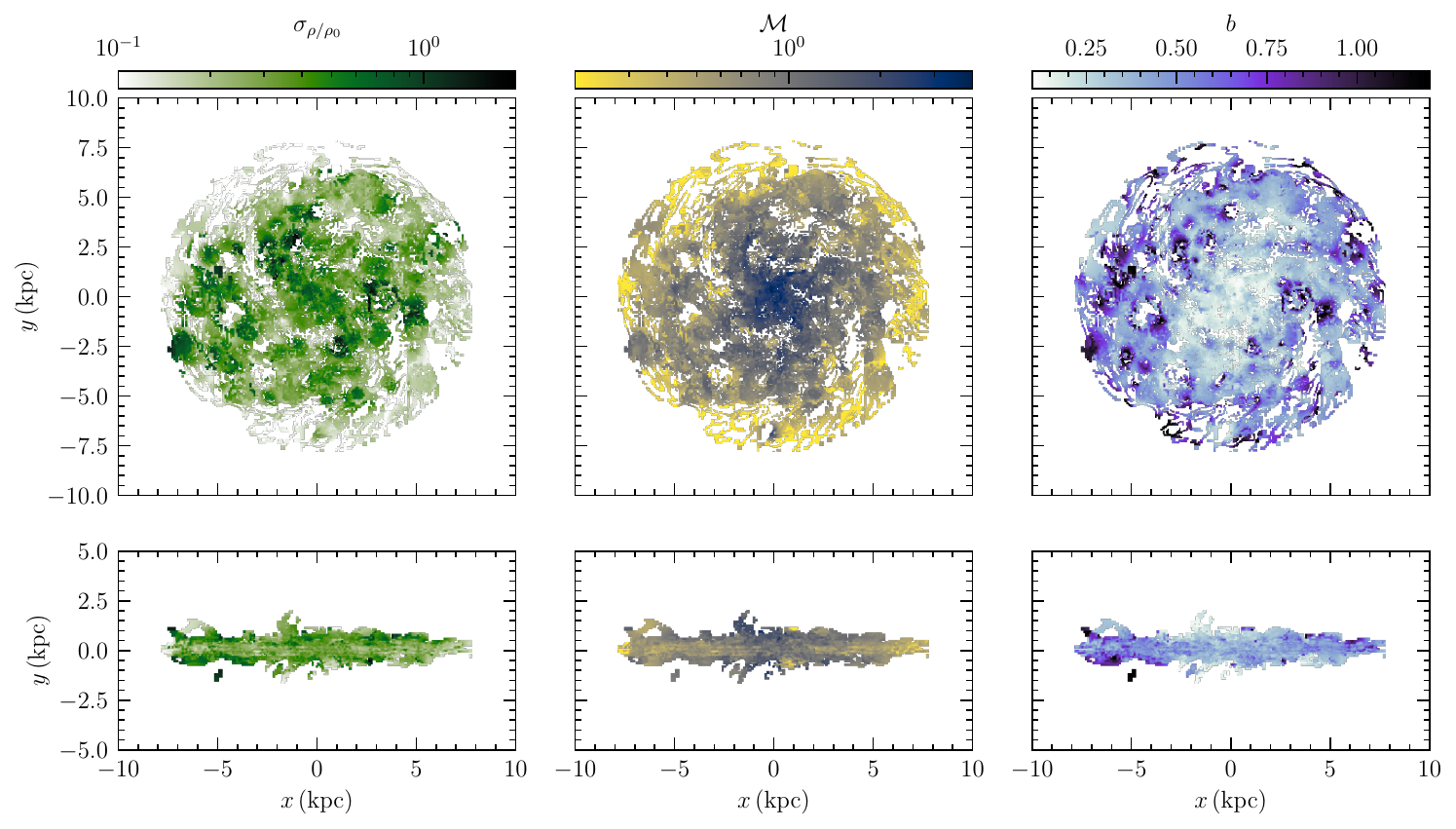}
    \caption{Volume-weighted projections of the density dispersion (left panel), turbulent Mach number (middle panel), and turbulence driving parameter $b$ (right panel). The density dispersion shows local enhancements, likely associated with supernova feedback and cloud-cloud collisions, but no clear radial trend is seen. In contrast, the Mach number increases towards the galactic centre, but remains overall in the transonic regime, as typical for WNM gas. The turbulence driving is more solenoidal ($b<0.4$) closer to the centre of the galaxy, for galactic radii $\lesssim 4\,\mathrm{kpc}$, and becomes $>0.4$ (dominance of compressive driving modes) at larger radii. In contrast to the radial dependencies, clear vertical trends are not seen.}
    \label{fig:b_map}
\end{figure*}

\begin{figure*}
    \centering
    \includegraphics[width=1\linewidth]{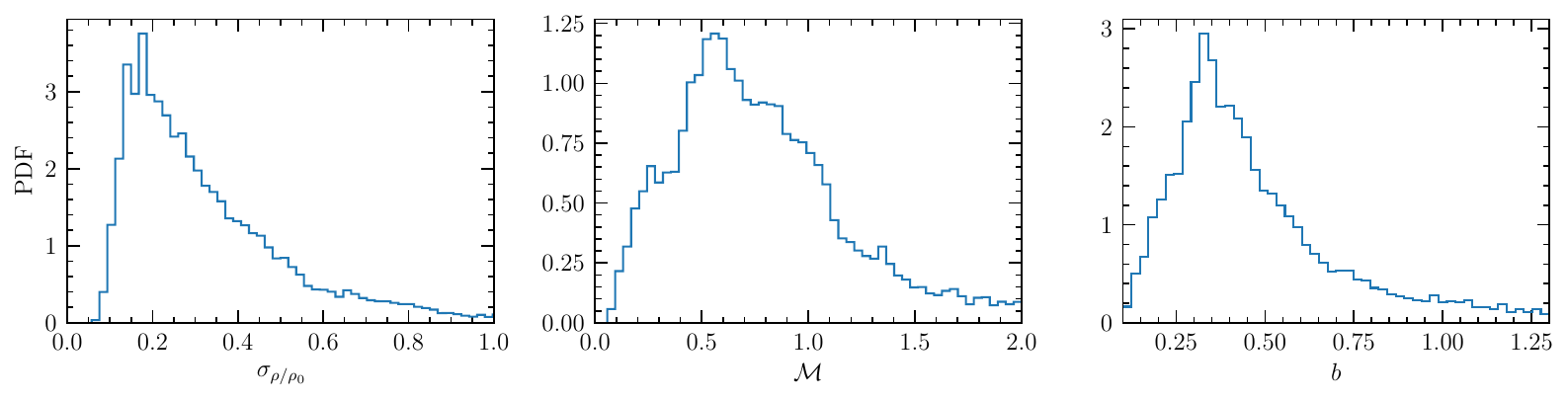}
    \caption{Volume-weighted PDFs of $\sigma_\mathrm{\rho/\rho_0}$ (left), $\mathcal{M}$ (middle), and $b$ (right). The Mach numbers indicate that the WNM is largely transonic. The turbulence driving is highly variable across the galaxy, ranging from solenoidal to compressive, with a median value of $\sim0.4$, i.e., consistent with the natural mixture of solenoidal and compressive driving ($b\approx0.4$).}
    \label{fig:turb_pdf}
\end{figure*}

Figure~\ref{fig:b_map} shows volume-weighted projections of the density dispersion, Mach number, and driving parameter, while Figure~\ref{fig:turb_pdf} presents their probability distributions. These quantities are highly variable across the galaxy, however, clear trends emerge: the Mach number increases towards the galactic centre, while the turbulence driving parameter decreases, indicating that the turbulence driving becomes increasingly solenoidal towards the galactic centre. In particular, the driving becomes predominantly solenoidal ($b<0.4$) within the inner $\lesssim 4\,\mathrm{kpc}$ of the galaxy. In contrast, all three quantities remain largely unchanged along the vertical direction. The Mach number spans $\approx 0.2$–$4$, placing most of the gas in the transonic regime, as expected for the WNM \citep[see also][]{McClureGriffithsEtAl2023,IsabellaEtAl2023}.

\begin{figure*}
    \centering
    \includegraphics[width=1\linewidth]{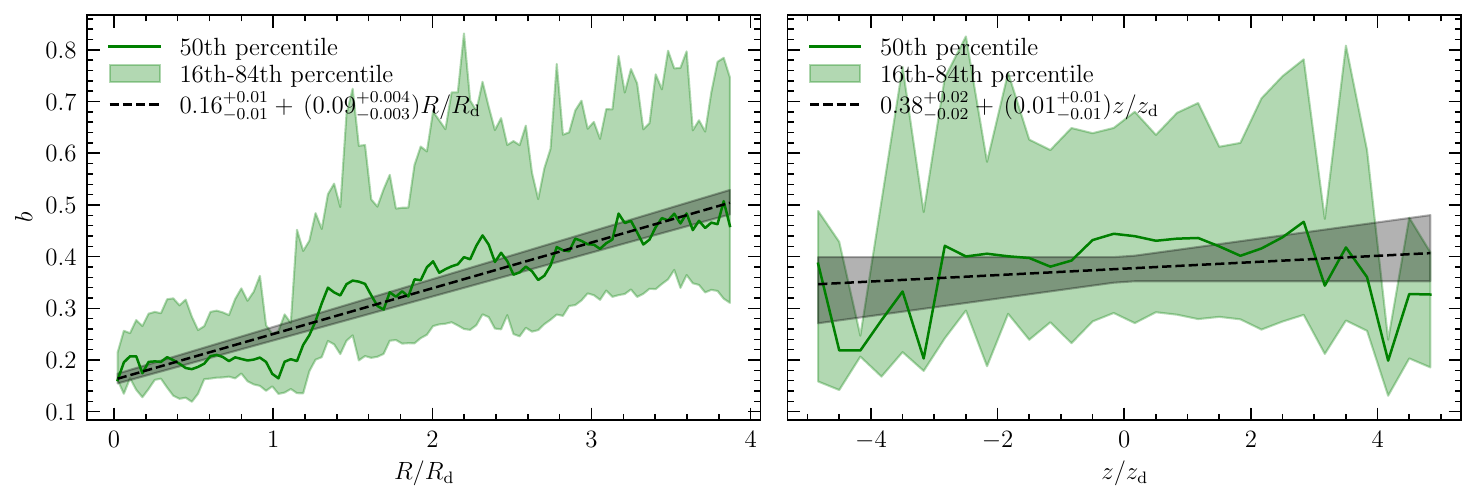}
    \caption{Variation of $b$ with galactic radius normalized by the initial scale length $R_d \approx 1.8$,\kpc (left panel) and disc height normalized by the initial stellar scale height $z_d \approx 0.2$\,\kpc (right panel). The driving mode of turbulence is more compressive as one moves radially away from the centre of the galaxy, however, it is largely unchanged along the $z$-axis. The dashed line shows a linear fit, with the fitted slope shown in the legend.}
    \label{fig:b_profile}
\end{figure*}

The high variability of the turbulence driving parameter may be the result of the local conditions and superposition of different driving mechanisms, such as supernova explosions, shear, etc. The map of $b$ in Figure~\ref{fig:b_profile} reinforces this picture: although the driving mode changes significantly from place to place, there is a systematic radial trend towards more compressive driving at larger galactocentric radii. Vertically, however, $b$ remains nearly constant. The left panel shows a roughly linear increase of $b$ with radius, well described by $b = 0.16^{+0.01}_{-0.01} + \left(0.090^{+0.004}_{-0.003}\right)R/R_\mathrm{d}$, whereas the right panel shows little variation along the $z$-axis, the fitted slope of $0.01\pm0.01$ being consistent with zero, with an average value of $b = 0.38 \pm 0.02$. Both the level of small-scale variability and the systematic radial increase of $b$ are consistent with the observational findings of \citet{MillerEtAl2026} in NCG1313 and NGC7793.

\subsection{Time evolution and correlation with shear}
\label{sec:correlation_with_shear}

Solenoidal driving is often attributed to shearing motions \citep{FederrathEtAl2016}. In a galaxy, these can arise from differential rotation. In order to quantify shear, we use the norm of the traceless symmetric part of the shear tensor. The shear tensor ($\mathcal{T}$) for a fluid is defined as
\begin{equation}
    \mathcal{T}_\mathrm{ij} = \frac{\partial v_\mathrm{i}}{\partial x_\mathrm{j}}\,,
    \label{eq:shear_tensor}
\end{equation}
where $v_\mathrm{i}$ is the $i$th component of the velocity. The traceless symmetric part of the shear tensor ($\mathcal{S}^\star$) is given by
\begin{equation}
    \mathcal{S^\star_\mathrm{ij}}=\frac{1}{2}(\mathcal{T}_\mathrm{ij}+\mathcal{T}_\mathrm{ji}) - \frac{1}{3}\delta_\mathrm{ij}\sum_{k=1}^{3}\mathcal{T}_\mathrm{kk}\,,
    \label{eq:traceless_symm_shear}
\end{equation}
where $\delta_\mathrm{ij}$ is the Kronecker delta tensor. The norm of $\mathcal{S}^\star$ is defined by 
\begin{equation}
    ||\mathcal{S}^\star||=\sqrt{\sum_\mathrm{i=1}^3\sum_\mathrm{j=1}^3\mathcal{S}^\star_\mathrm{ij}\mathcal{S}^\star_\mathrm{ij}}\,.
    \label{eq:shear_norm}
\end{equation}

For a consistent comparison, we Gaussian-average the norm of $\mathcal{S}^\star$ over the same length scales as the turbulent Mach number and the density dispersion. More rigorously, for $\mathcal{U}$ the set of all cells in our simulation, and $\mathcal{P}$ the set of all WNM points, we define
\begin{equation}
    \mathcal{F}_p=\{a\in\mathcal{U}:||p-a||^2\le\alpha||D(p)||^2\}\,,
\end{equation}
where $p\in\mathcal{P}$ and $\alpha=6.5$. The average value of $||\mathcal{S}^\star||$ is then given by $\langle||\mathcal{S}^\star||\rangle_{(\mathcal{F}_{p};\gamma)}$, where $\gamma=5||D(p)||$ (see Eq.~\ref{eq:unstructured_gaussian_mean}).

\begin{figure}
    \centering
    \includegraphics[width=1\linewidth]{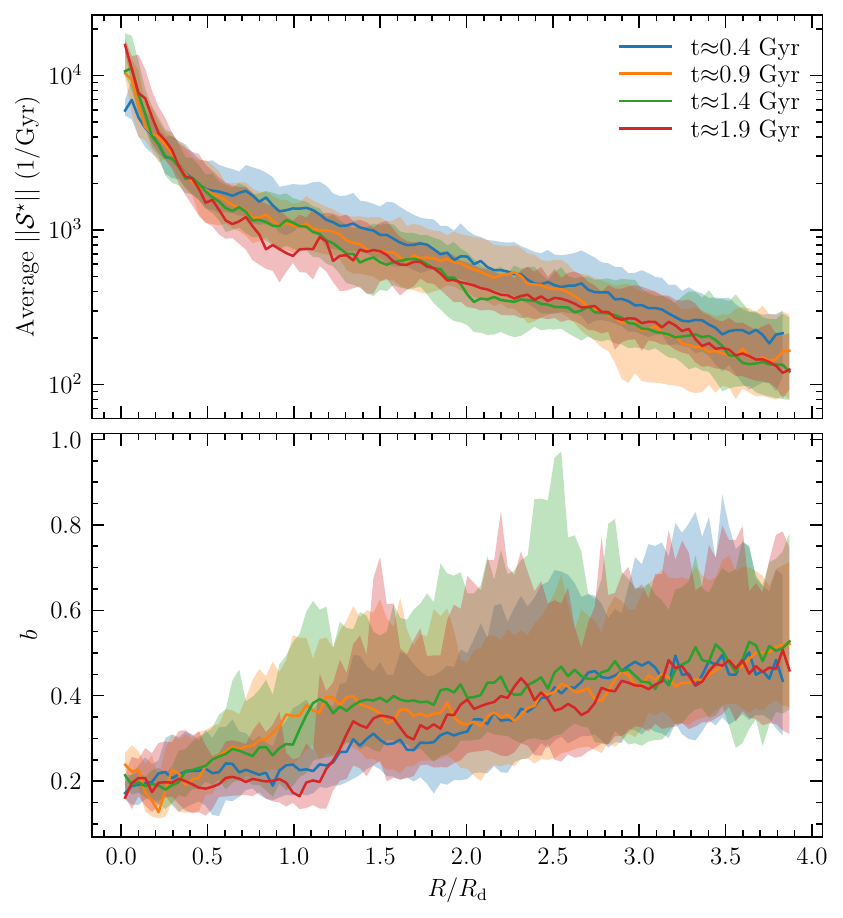}
    \caption{Spatially averaged norm (as described in Section~\ref{sec:correlation_with_shear}) of the traceless symmetric shear tensor (top panel) and the turbulence driving parameter $b$ (bottom panel) as a function of galactic radius normalized by the initial scale length, at four times ($t\approx0.4$, $0.9$, $1.4$, and $1.9$\,Gyr) during the evolution of the galaxy. High-shear environments close to the galactic centre are associated with solenoidal driving of turbulence.}
    \label{fig:shear_profile}
\end{figure}

\begin{figure*}
    \centering
    \includegraphics[width=1\linewidth]{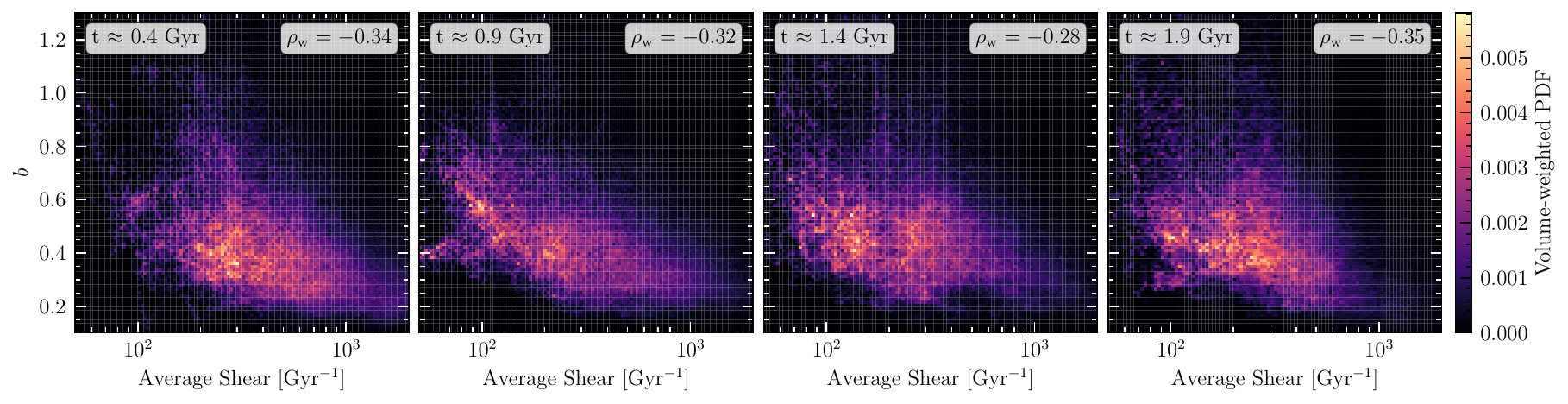}
    \caption{Correlation PDF of $b$ versus the average norm of the traceless symmetric shear tensor at the four different times (from left to right) shown in Fig.~\ref{fig:shear_profile}. The inset on the top right shows the volume-weighted Spearman rank correlation coefficient. There is a weak anti-correlation between $b$ and shear, however, the anti-correlation is stronger for low-$b$ values. This is attributed to solenoidal driving being associated with high-shear environments.}
    \label{fig:correl}
\end{figure*}

We plot the average shear profile and the $b$ profile at four different times ($\approx\!0.4, 0.9, 1.4$, and $1.9\,$Gyr) in Figure~\ref{fig:shear_profile}. The profiles are largely identical at different times during the galaxy evolution. The shear increases as one moves closer towards the galactic centre due to an increase in the strength of differential rotation. This is accompanied by a decrease in the turbulence driving parameter $b$, i.e., towards more solenoidal driving. We also perform a cell-by-cell comparison of shear with $b$ in Figure~\ref{fig:correl}. The inset on the top right shows the volume-weighted Spearman rank correlation coefficient ($\rho_\mathrm{w}$), which ranges from $-0.28$ to $-0.35$ across the four epochs. There is a weak, but significant anti-correlation between shear and $b$ and it is the tail of the distribution (corresponding to low $b$) that has a stronger anti-correlation with shear. This confirms that solenoidal driving is associated with high-shear environments. The lack of correlation for higher values of $b$ can be attributed to other local drivers of turbulence, like, supernovae.

\section{Conclusions}
\label{sec:conclusions}

Much of our quantitative understanding of astrophysical turbulence rests on idealised, periodic-box simulations in which turbulence is sustained by a stochastic forcing field of prescribed geometry. The turbulence driving parameter $b$ is the quantity that connects those controlled experiments to real, inhomogeneous systems, since it sets the width of the density PDF and thereby the star formation rate \citep{FederrathKlessen2012}. In this work we have built that bridge from the simulation side, by measuring $b$ throughout an entire simulated galactic disc rather than in a single cloud or patch. Our main results are as follows.

\begin{enumerate}
    \item We have generalised the kernel-based algorithm of \citet{IsabellaEtAl2023} for measuring the turbulence driving parameter to AMR simulations. The key modification is that the roving and smoothing kernels adapt to the local cell size, and that all averages are volume-weighted and renormalised over the cells that satisfy the phase criterion (Eq.~\ref{eq:unstructured_gaussian_mean}). This allows the background subtraction to be performed consistently across refinement levels and in the presence of gaps in the kernel. Applying the algorithm to an LMC-mass galaxy simulated with {\sc Nexus} \citep{GarciaEtAl2024, bla24a}, we recover a nearly log-normal density PDF and nearly Gaussian velocity PDFs over $\gtrsim 90\%$ of the data after background removal (Fig.~\ref{fig:den_vel_hist}), confirming that the procedure isolates the turbulent component of the flow.

    \item The driving parameter varies strongly across the disc, spanning the full range from solenoidal to compressive driving, with a volume-weighted median of $b\simeq0.4$, i.e., consistent with the natural mixture of solenoidal and compressive driving \citep[$b\approx0.4$;][]{FederrathEtAl2010}, rather than being systematically biased towards either extreme (Fig.~\ref{fig:turb_pdf}). The turbulent Mach number of the WNM lies in the range $\mathcal{M}\approx0.2$--$4$, i.e.\ the warm phase is largely transonic, consistent with \citet{IsabellaEtAl2023}.

    \item Superimposed on the scatter in $b$ is a systematic radial trend: the driving is predominantly solenoidal ($b<0.4$) within the inner $\lesssim4\,\mathrm{kpc}$ and becomes progressively more compressive outwards, increasing approximately linearly with galactocentric radius as $b = 0.16^{+0.01}_{-0.01} + \left(0.090^{+0.004}_{-0.003}\right)R/R_\mathrm{d}$ (Fig.~\ref{fig:b_profile}). By contrast, $b$ is nearly independent of height above or below the mid-plane: the corresponding vertical fit has a slope of $0.01\pm0.01$, consistent with zero, and an average of $b = 0.38 \pm 0.02$. Both the large cell-to-cell variability and the systematic radial increase agree with the observational measurements of \citet{MillerEtAl2026}.

    \item The shear, quantified by the norm of the traceless symmetric part of the shear tensor and averaged over the same scales as $b$, rises towards the galactic centre as differential rotation strengthens, mirroring the decline of $b$ in the same region (Fig.~\ref{fig:shear_profile}). A cell-by-cell comparison yields a weak volume-weighted anti-correlation between $b$ and shear, with a Spearman rank correlation coefficient of $\rho_\mathrm{w}\approx-0.28$ to $-0.35$ across the epochs analysed (Fig.~\ref{fig:correl}). The anti-correlation is carried mainly by the low-$b$ tail, confirming that solenoidal driving is associated with high-shear environments, while the weak correlation at high $b$ suggests that other, more local mechanisms --- most plausibly supernova feedback --- dominate the compressive driving elsewhere in the disc.

    \item All of these trends --- the radial profile of $b$, the radial profile of the shear, and the weak anti-correlation between them --- are essentially unchanged at the four epochs we examined, spanning the evolution of the galaxy. The driving mode of turbulence therefore appears to be a robust, quasi-steady property of the disc, set by its large-scale structure rather than by transient events, which only modulate the local turbulence driving in a stochastic manner.
\end{enumerate}

Taken together, these results caution against adopting a single, universal value of $b$ in sub-grid models of star formation. The driving parameter is a local quantity that varies systematically with galactocentric radius, and a model calibrated on the solenoidally driven inner disc will not describe the more compressively driven outskirts. At the same time, the weak correlation with shear indicates that $b$ cannot simply be predicted from the large-scale dynamics alone.

In this work, we have restricted the analysis to the WNM in order to keep the sound speed and hence the Mach number well defined. In future works, we plan to extend the measurement to other phases of the ISM. Our analysis is also limited to a single galaxy model. A comparative study across the {\sc Nexus} suite --- varying gas fraction, feedback prescription, galaxy mass etc.~--- would establish which of the trends reported here are generic and which are specific to an LMC-like system. Finally, disentangling the individual contributions of shear, supernova feedback and gravitational instability to the local driving mode remains an open problem.

\section*{Acknowledgements}
We thank Amit Seta and Mark Krumholz for helpful discussions. C.F.~acknowledges funding provided by the Australian Research Council (Discovery Projects DP230102280 and DP250101526), and the Australia-Germany Joint Research Cooperation Scheme (UA-DAAD). TTG acknowledges financial support from the Australian Research Council (ARC) through Australian Laureate Fellowships awarded to JBH (FL140100278) and TRB (FL220100117), for partial funding from Lund University, and from the James Arthur Pollock memorial fund awarded to the School of Physics, University of Sydney. JBH also acknowledges support from the ARC grant DP220103384. OA acknowledges support from the Knut and Alice Wallenberg Foundation, the Swedish Research Council (grant 2025-04892), the Swedish National Space Agency (SNSA grants 2023-00164 and 2025-00405), the LMK foundation, and eSSENCE, a Swedish strategic research programme in e-Science.
The authors acknowledge LUNARC, the Centre for Scientific and Technical Computing at Lund University, for providing computational resources on the COSMOS cluster.
We further acknowledge high-performance computing resources provided by the Leibniz Rechenzentrum and the Gauss Centre for Supercomputing (grants~pr32lo, pr48pi and GCS Large-scale project~10391), the Australian National Computational Infrastructure (grant~ek9) and the Pawsey Supercomputing Centre (project~pawsey0810) in the framework of the National Computational Merit Allocation Scheme and the ANU Merit Allocation Scheme.
We acknowledge the use of large language model (LLM) based assistants, namely GitHub Copilot, ChatGPT and Claude, for code optimisation, for generating draft text for the Conclusions and Appendix~\ref{app:kernel_size}, and for refining the text throughout the manuscript. All AI-generated content was reviewed, verified and edited by the authors, who take full responsibility for the content of this paper.


\section*{Data Availability}
The simulation data ($\sim6\,\mathrm{GB}$) underlying this article will be shared on reasonable request to the authors. The implementation of our algorithm used for calculating the turbulence driving parameter in this work is available at \url{https://github.com/James471/TurbulenceScripts/tree/10846b0f0204cb87b8fc975f0d564bee2a81d578}.



\bibliographystyle{mnras}
\bibliography{references} 




\appendix

\section{Choice of kernel size}
\label{app:kernel_size}

The roving kernel sets the scale on which turbulence is measured, so it is important to establish how sensitive our results are to it. Here we quantify that sensitivity, and motivate the value $\lambda=5$ adopted in Section~\ref{sec:method}.

\begin{figure}
    \centering
    \includegraphics[width=1\linewidth]{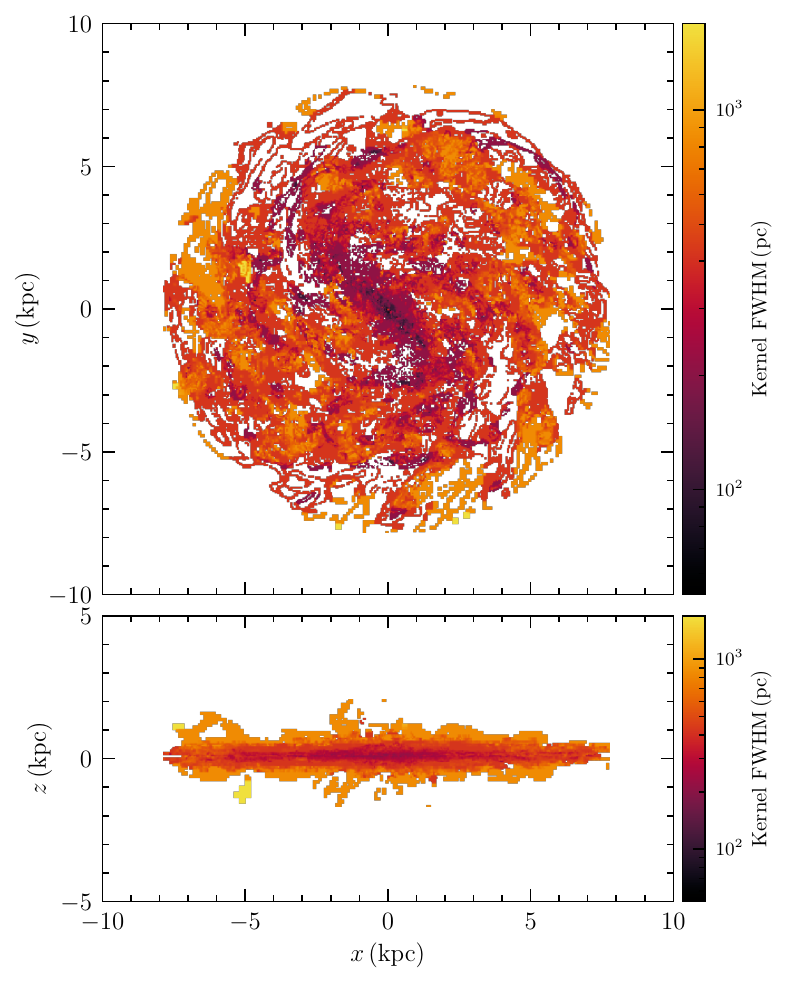}
    \caption{Projected map of the roving-kernel FWHM, $\lambda\|D(p)\|$ with $\lambda=5$, along the $z$-axis (top) and the $y$-axis (bottom). The kernel adapts to the local refinement level, contracting to a few tens of parsecs in the dense, well-resolved gas near the mid-plane and expanding to several hundred parsecs in the diffuse outer disc.}
    \label{fig:kernel_map}
\end{figure}

Because the grid is adaptive, the kernel is not a fixed physical size: its FWHM, $\lambda\|D(p)\|$, follows the local cell size and therefore varies by more than an order of magnitude across the disc. Fig.~\ref{fig:kernel_map} maps the projected kernel FWHM for $\lambda=5$. In the dense, highly refined gas near the mid-plane the kernel is a few tens of parsecs across, while in the diffuse outskirts and at large heights it grows to several hundred parsecs. The kernel thus automatically probes smaller scales where the simulation resolves them, which is the principal advantage of the adaptive formulation over the fixed-size kernel of \citet{IsabellaEtAl2023}.

The choice of $\lambda$ is a compromise. It must be large enough that each kernel contains sufficient cells for the dispersions in Eqs.~\eqref{eq:sigma_rho}--\eqref{eq:mach_calculation} to be well sampled, and small enough to resolve the spatial variation of the turbulence quantities across the galaxy. To test the impact of this choice we repeated the full calculation for $\lambda=3,4,5,6$ and $7$, keeping $\beta$ and the ratio $\gamma/\lambda$ fixed so that the smoothing and weighting scales track the kernel.

\begin{figure}
    \centering
    \includegraphics[width=1\linewidth]{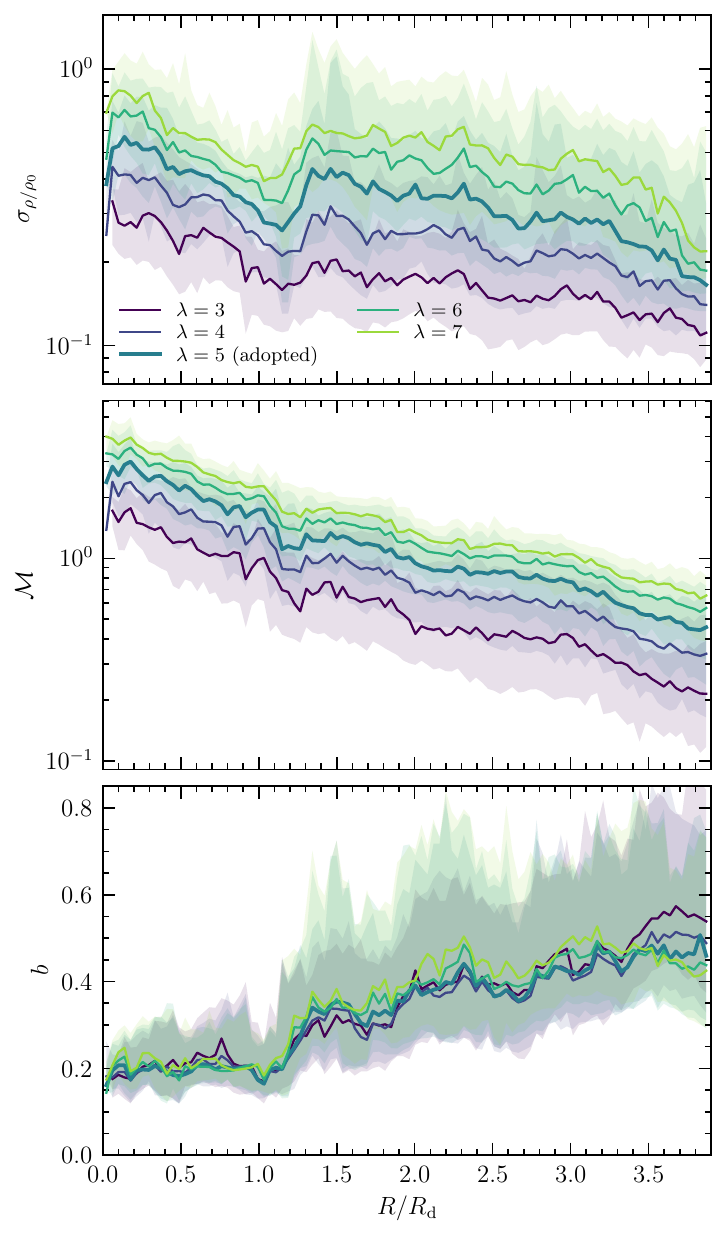}
    \caption{Volume-weighted radial profiles of the density dispersion (top), turbulent Mach number (middle) and driving parameter (bottom), for roving-kernel FWHM factors $\lambda=3$--$7$. Shaded bands show the 16th--84th percentile range. While $\sigma_{\rho/\rho_0}$ and $\mathcal{M}$ both depend strongly on the kernel size, their ratio $b$ is far less sensitive, and the radial trend is recovered for every value of $\lambda$.}
    \label{fig:kernel_size_dependence}
\end{figure}

Fig.~\ref{fig:kernel_size_dependence} shows the resulting radial profiles of $\sigma_{\rho/\rho_0}$, $\mathcal{M}$ and $b$. As expected for turbulent fields, the first two are strongly scale dependent: both increase systematically with $\lambda$, since a larger kernel admits velocity and density fluctuations on larger scales. Their ratio, however, is far more stable. To quantify this we compare the five profiles of each quantity bin by bin: at a given radius we take the difference between the largest and smallest of the five values and divide it by their mean, and we then report the median of this fractional spread. The spread across $\lambda=3$--$7$ is 103\% in $\sigma_{\rho/\rho_0}$ and 86\% in $\mathcal{M}$, but only 15\% in $b$. Crucially, the radial trend itself --- solenoidal driving in the inner disc giving way to more compressive driving outwards --- is present at every $\lambda$ we tested, so none of the conclusions of Section~\ref{sec:results} depend on this choice.

This behaviour is the same as that reported by \citet{IsabellaEtAl2023}, and has a straightforward interpretation: $\sigma_{\rho/\rho_0}$ and $\mathcal{M}$ both change with the scale on which they are measured, in such a way that the change largely cancels in their ratio. The driving parameter is therefore a comparatively robust diagnostic, provided the kernel size is not varied by factors of several, in which case turbulence on a genuinely different scale is being probed.

We adopt $\lambda=5$ as the smallest value for which the kernels are well populated: it yields a median of 292 cells per roving kernel, comfortably above the threshold of 30 imposed in Section~\ref{sec:method}, while keeping the kernel small enough to resolve structure on scales well below the disc scale length.

\bsp	
\label{lastpage}
\end{document}